\relax
\documentclass[prl,aps]{revtex4}

\usepackage{graphicx}

\begin{document}
\title{ Intersite elastic coupling and invar effect}

\author{D. I. Khomskii   $^1$ and  F. V. Kusmartsev $^{2}$}
\date{ 20.01.2004}

\address{{$^1$II Physikalisches Institut}, {Universit$ \ddot a$t 
zu K$\ddot o$ln, }{Z$\ddot u$lpicher Str. 77}
\centerline{50937 K$\ddot o$ln, Germany}
\centerline{$^2$Department of Physics, Loughborough University, LE11 3TU, UK}}

\begin{abstract}
The invar phenomenon  (very small thermal expansion in some iron
alloys or compounds) is usually explained by the thermally-induced
transitions between different spin states of $Fe$, having
different atomic volumes.  We consider these processes taking
into account elastic interaction between $Fe$ atoms in different
spin states. Inclusion of these interactions  explains why thermal 
expansion may be close to zero in a broad temperature interval and 
thus gives rise to the invar effect.

\end{abstract}

\maketitle

\bigskip
\bigskip
  Invar behaviour -- the absence of the dependence of the lattice 
parameter on temperature
  in $Fe-Ni$ alloys in certain concentration range, was discovered in 
1897 [1], and similar
  behaviour was found later in certain other systems - e.g. in ordered 
and disordered $Fe_3\  Pt$
  and $Fe_3\  Pd.$ [2].  The most plausible explanation of this 
phenomenon was suggested by Weiss [3]
  who postulated the existence of two states of iron, close in energy: 
the ground state with the high
  spin, or the high-moment $(HM)$ state with large specific volume, 
and the low-lying excited state
  with low spin or low moment $(LM)$, having smaller atomic radius or 
specific volume.  According
  to this picture, thermal excitation of the $LM$ low-volume states 
causes lattice contraction,
  which counteracts and may cancel the usual positive thermal 
expansion.  Although there is yet no
  definite proof of the existence of such two states in invar alloys, 
many experimented facts are
  naturally explained in this picture [2,4,5].  The existence of 
almost degenerate states with
  different moments and different specific volumes is also 
corroborated by the detailed  band-structure  calculations [6,7].

Recent neutron scattering studies [8] have confirmed the importance 
of magnetoelastic coupling
for the invar effect $-$ apparently not the usual coupling present in 
magnetic materials with a
given spin of the ions, but of the coupling with mutliplet 
excitations, e.g. $HM-LM$ excitations
iron.  Although many particular details are still not clear, all 
these results confirm the general
validity of the Weiss two-state model.

This simple explanation of the invar effect is very appealing. 
However, one problem in this
explanation becomes immediately apparent. The conventional thermal 
expansion is usually
more or less linear in temperature
$$
a(T)=a_0+\alpha_{0} T  , \eqno{(1)}
$$
where $a(T)$ is the lattice parameter at a temperature $T$ and 
$\alpha_0$ is the conventional
thermal expansion coefficient. On the other hand the thermal 
population of the low-spin state with smaller radius
in simplest case of two well-defined $LM$ and $HM$ states would be 
exponential in temperature:
$$
a(T)=a_0-c  \exp({-{\Delta\over T}})  , \eqno{(2)}
$$
where  $\triangle=E_{L}-E_{H}$ is the excitation energy of the $LM$ 
state, and $a_0=a_H$;
$c=(a_{H}-a_{L})/2$;
$a_{H/L}$ are the ionic radii of corresponding spin states.  Thus, 
the question arises, how can
one compensate in a reasonably broad temperature interval the normal 
positive thermal expansion (1),
linear in $T$, by the extra negative contribution (2) which depends 
on the temperature {\it exponentially}.

In this paper we suggest the simple mechanism which should always 
exist in real materials and which
helps to resolve this paradox.  When one discusses the coupling of 
the electronic excitations
(here HM - LM excitation) to the lattice, this interaction, besides 
coupling the electronic static
with local deformation, usually leads also to an effective 
interaction {\it between different sites} (somewhat similar interaction was also 
taken into account  by Gr\"uner et al. [9] in their Monte Carlo numerical simulations).
One can easily show that if we consider predominantly a coupling to 
the short-range (or optical)
vibrations, this intersite interaction will be essentially of 
antiferro type [10]:  If we transform one site
from a $HM$ to a  $LM$ state with smaller volume, it would be 
favourable to have close to this
small-volume $LM$ ion the larger, i.e. $HM$ ions.  This interaction 
will modify the temperature
dependence of the occupation of different spin-states, and, 
consequently, will change the extra
contribution to thermal expansion, effectively stretching the 
exponential temperature dependence (2).
This would help to explain the almost full compensation of two 
mechanisms of thermal expansion --
the usual one (1) and the additional stretched contribution, giving 
finally the invar effect in a
rather broad temperature interval.

One can describe this situation introducing the pseudospin operators, 
which describe
two spin states, so that the state $\tau^z_i=+{1\over 2}$ corresponds 
to the $HM$ state of an ion
$i$ and $\tau^z_i=-{1\over 2}$ to a $LM$ state of it.  The fact that 
these states have different
ions radii (or atomic volumes) gives rise to a coupling of these 
states to the lattice, which
classically can be written as:

$$
H=-g\tau^z_i\big(v_i-v_0\big)+{B\over 2}\big(v_i-v_0\big)^2-\Delta 
\tau^z_i\eqno{(3)}
$$

Here $v_0$ is an average volume,  $v_0={1\over 2}\big(v_L+v_H)$,  where $v_L$ and $v_H$ are the corresponding atomic volumes of the respectively $LM$ and $HM$ states,
and $g={B}\big(v_H-v_L\big)$ is the effective coupling constant.  By 
minimizing the average
energy $E=<H>$ with respect to volume, we can indeed see that
$$
v_i = v_0+{g\over B}\tau^z_i   , \eqno{(4)}
$$
which, with our choice of $v_0$ and $g$, reproduce the correct results,
$
v_i\big(\tau={1\over 2}\big)= v_H,\ \ v_i \big(\tau=-{1\over 
2}\big)=v_L
$. We included in the Hamiltonian (3) also the term with 
the ``magnetic field'', $-\Delta \tau^z_i$,
which describes the initial splitting of the $HM$ and $LM$ states: 
$\triangle=E_L-E_H$.

The model (3) describes only the single-site effects.  But when one 
takes into account the coupling
of local distortions around different sites (giving rise to the 
dispersion of phonons), one would get,
besides these on-site effects, also an intersite interaction.  If one 
rewrites the model (3) including
the phonons dispersion,

$$
H=\sum_{i,k} 
\tilde 
g_{ik} \tau^z_i\big(  b^{\dagger}_k+ 
b_k\big)+\sum_k \omega_k  b^{\dagger}_k  b_k - \Delta \sum_i 
\tau_i^z, \eqno{(5)}
$$
where  $\tilde g_{ik}=\tilde g_k e^{ikRi}$, one can in the usual way 
exclude the phonons by
canonical transformation and obtain the effective pseudospin 
Hamiltonian, see e.g. [11]:

$$
H_{eff}=\sum_{ij} {\cal J}_{ij}  \tau^z_i \tau^z_j - \Delta \sum_i 
\tau^z_i \eqno{(6)}
$$

$$
{\cal J}_{ij}=-\sum_k\ e^{ik(R_i-R_j)} { \tilde g_{k}^2 \over \omega_k}  
$$

The effective sign of an intersite interaction depends on the 
detailed $k$-dependence of the
spin-phonon matrix element $\tilde g_k$, on the phonon dispersion 
$\omega_k$ and on the type of the
lattice.  One can easily show that the coupling via short-wavelength 
phonons leads to a nearest-neighbor
repulsion ${\cal J} >1$, i.e. to an antiferromagnetic interaction 
between pseudospins $\tau$, in accordance with
the qualitative considerations presented above (the large $HM$ state 
$\tau^z_1=+{1\over 2}$ would
prefer to have nearby the low-volume $LM$ sites, $\tau^z_j=-{1\over 
2}$).  Longer range interactions
may in general have different sign [12], but usually the $nn$ 
interactions dominate, and this is what
we will assume further on.

With this assumption we can reduce our model to an antiferromagnetic 
Ising model with $nn$ coupling
${\cal J}$ in a parallel field.  For invar systems, the parameters of 
the model should be chosen such that the ground state corresponds to 
the $HM$ state, i.e. all $\tau^z_1=+{1\over 2}$, which requires 
$\Delta > {\cal J}$.  In this case the standard mean-field equation 
for the total (not sublattice!) magnetization takes the form:

$$
\tau=<\tau>={1\over 2}\ th{\Delta -2{\cal J} z\tau\over 2T}\eqno{(7)}
$$
($z$ is the number of nearest neighbours), from which we can 
determine the temperature dependence
of $\tau$ and consequently, according to (4), of the average volume 
of our system,
$$v(T)=v_0+{g\over B}\tau (T)\eqno{(8)}$$
It is convenient  to rewrite Eq. (7) as
$$\tau={1\over 2}th{\tilde\Delta+2{\cal J}z\left({1\over 
2}-\tau\right)\over 2T}   , \eqno{(9)}$$
where $\tilde\Delta =\Delta +2{\cal J}z\tau(0)=\Delta + {\cal J}z$ is 
the renormalized
initial $(T=0)$ splitting of the $LM$ and $HM$ states. If we would 
take this splitting to be constant
(i.e. if we ignore the second term in the argument of Eq. (9)),  we 
would get the conventional temperature
dependence of $\tau$ (Brillouin function)
and, consequently, of the lattice parameter and of the thermal expansion,
which at low temperature would be exponential in temperature:

$$\tau (T)={1\over 2}- \exp \bigg( - {\tilde\Delta\over T} 
\bigg)  , \eqno{(10)}
$$

$$
v(T)=v_H- {(v_H-v_L)\over 2}    \exp \left(-{\tilde\Delta\over 
T}\right) , 
$$
cf. (2) (here $v(T=0)=v_H)$.  This is what one would naively get in the
standard Weiss model, which ignores the intersite interaction.  As 
discussed above, we have then
the problem, how this exponental contribution can compensate the 
usual linear positive thermal
expansion in a broad temperature interval.

The analysis of equation (9) shows that when we include the intersite 
interaction, it leads,
besides the renormalisation of the initial splitting of $LH$ and $HM$ 
states, to the modification
of the temperatures dependence of $\tau$ and, correspondingly, of the 
lattice parameters.  This
is shown in Fig.1, in which we present the results of the calculations for 
representative values of parameters $\Delta$ = 550 K, ${\cal J}$z = 440 K. 
The dotted line is the dependence 
of $\tau(T)$ (or of an extra contribution to the
volume $v(T)$) ignoring the intersite elastic inteaction (the term 
with ${\cal J}$ in equation (9)),  and the solid line -- with this 
interaction taken into account . By thin line  we qualitatively show 
the conventional positive thermal expansion which behaves as $\sim 
T^4$ at low temperatures and goes over to linear dependence for 
higher T. We see indeed that, whereas without intersite interaction 
(${\cal J}=0$), $\tau$ changes with
temperatures rather steeply (initially as  $ {1\over 2} - \exp \left( 
-{\tilde\Delta \over T}\right) $ ),  with non-zero intersite coupling 
${\cal J}$ this dependence becomes much smoother.This is easy to 
understand:  if indeed there exists a repulsion between similar spin 
states (antiferromagnetic coupling in equation (6)), then the thermal 
excitations of certain amount
of $LM$ states hinder corresponding transitions on neighbouring 
sites, so that as a result the
average excitation energy $\tilde\Delta +{\cal J}z\left({1\over 
2}-\tau(T)\right)$ would gradually
increase with temperature, making the extra negative contribution to 
lattice parameter more smooth.
If we now add to this term the usual positive thermal expansion 
(qualitatively shown in Fig.1 by
  thin line), we indeed see that with ${\cal J}\neq 0$ one can get a 
better cancellation of the
  normal and anomalous contributions to thermal expansion (although 
this cancellation is not exact).

\vskip 6truept
\hskip.01truein
\includegraphics{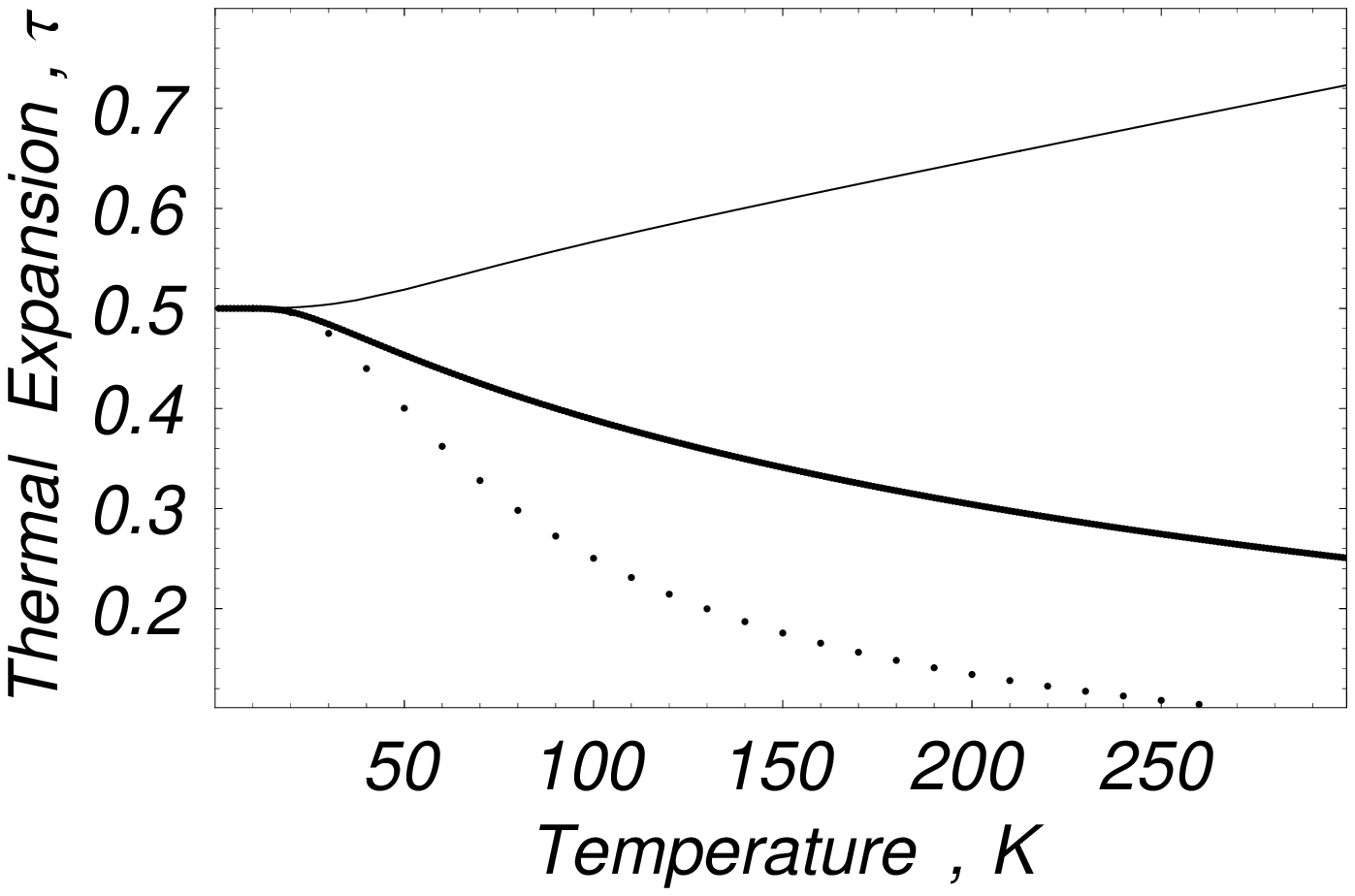}

\vskip 6truept 
\noindent
{\bf Fig.1} The extra negative thermal expansion(see Eq.(9)) without 
intersite elastic interaction (dotted line) and with this interaction 
taken into account (solid line). Thin line is a conventional positive 
thermal expansion.

\bigskip

This is the main conclusion of the present paper.  We used the fact 
that the elastic interaction
between different spin (and volume) states of $Fe$ in invar alloys is 
always present.  We show that
its inclusion helps to resolve some of the problems inherent to the 
two-state (Weiss) model
traditionally used in this field. The essence of our results is that 
due to this interaction the effective energy separating low moment/small 
volume and high moment/large volume states in invar alloys becomes  temperature-dependent. This modifies the temperature dependence 
of the thermal expansion and finally guarantees the invar behaviour in a 
broad temperature interval.

Extra consequences of our treatment are, first, that due to this effect 
the energy separation of these two states becomes dependent on the local 
coordination (occupation of neighbouring sites); this can hinder the direct 
observation of these two-level-systems e.g. by the neutron scattering. On the other 
hand, there should appear certain correlation in the 
occupation of different magnetic states;  this  effect should be observable
experimentally.  This could even lead to the formation of some textures in the invar samples. 

Many of the problems in this field still remain open.  One of them is 
the role of magnetic ordering
for the invar phenomenon, which was not included in the present 
treatment.  Another problem is the
account of the metallic nature of most of the invar systems. 
Nevertheless, even in this simplified
form the model considered above, with the inclusion of the intersite 
elastic interactions, can explain
the main features of the invar systems, and these interactions 
definitely have to be taken into account in the full
theory of the invar effect.

\bigskip

We are grateful to M. Abd-Elmeguid, K. Neumann and K.R.A. Ziebeck for 
useful discussions.  This work
was supported by the German Physical Society via SFB 608  and by the 
Leverhulme Trust.

\vfil\eject

\end{document}